\documentclass{article}

\usepackage[preprint]{nips_2018}

\usepackage[utf8]{inputenc} 
\usepackage[T1]{fontenc}    
\usepackage{hyperref}       
\usepackage{url}            
\usepackage{booktabs}       
\usepackage{amsfonts}       
\usepackage{nicefrac}       
\usepackage{microtype}      
\usepackage{color}
\usepackage{float}

\usepackage{graphicx}

\title{Dual-track Music Generation using Deep Learning}

\author{
  Sudi Lyu \\
  School of Computer Science\\
  Carnegie Mellon University\\ 
  Pittsburgh, PA 15213 \\
  \texttt{sudil@andrew.cmu.edu}\\
  \And
  Anxiang Zhang \\
  School of Computer Science\\
  Carnegie Mellon University\\ 
  Pittsburgh, PA 15213 \\
  \texttt{anxiangz@andrew.cmu.edu}\\
  \And
  Rong Song \\
  School of Computer Science\\
  Carnegie Mellon University\\ 
  Pittsburgh, PA 15213 \\
  \texttt{rongsong@andrew.cmu.edu}\\
}

\begin{document}

\maketitle
\begin{abstract}
  Music generation is always interesting in a sense that there is no formalized recipe. In this work, we propose a novel dual-track architecture for generating classical piano music, which is able to model the inter-dependency of left-hand and right-hand piano music. Particularly, we experimented with a lot of different models of neural network as well as different representations of music, and the results show that our proposed model outperforms all other tested methods. Besides, we deployed some special policies for model training and generation, which contributed to the model performance remarkably. Finally, under two evaluation methods, we compared our models with the MuseGAN project and true music.
\end{abstract}

\section{Introduction}
Music generation is quite a challenging but always exciting task since there can be a great number of possible combinations for musical notes and lots of variations like harmony. For this task, generative methods are usually deployed. For instance, the Hidden Markov Model can be used to generate a music note series. Particularly, recent advancements in deep learning methods have brought novel algorithms for this problem. The time-series model is usually deployed, such as RNN, LSTM, Encoder-Decoder, and attention mechanism. In this work, we mainly focus on music generation methods based on deep learning.

Compared with other generative tasks such as generating images or text, music generation has some notable differences. One most obvious thing is that information in music has a lot to do with time. So it attaches much importance to find a proper representation between time relations in the generating process. Besides, music is composed by some specific instruments. What's more, musical notes may form chords, and further constitute different parts of music such as melodies, low end, etc. Inspired by the field knowledge in playing the piano, we propose a new architecture that could benefit from the two parts of MIDI-based music, which are melodies and the low-end part. 

To build a strong generator as part of our proposed model. Specifically, we employ teaching force rate and randomization to optimize our model. As a result, the optimized model is predominant to all other tested methods and is expected to be even comparable with some of the state-of-the-art methods when deployed under large-scale datasets. 

At last, we conduct an abundant of experiments to verify the performance of different time-series model. We did 1) manual evaluation and 2) quantitative evaluation. We also compared our results to Google's MuseGAN model. 


In summary, our main contributions are as follows:
\begin{itemize}
    \item We propose a novel architecture for MIDI-based music generation.
    
    \item We explore different combinations of LSTM, CNN, encoder-decoder and attention model, and compare their performance for music generation.
    
    \item We employ some special training techniques to improve the performance of our models.
\end{itemize}

\section{Related Works}
\label{gen_inst}
According to the types of music representation, music generation approaches can be categorized as waveform-based methods and MIDI-based methods. In this work, we work on both but pay more attention to the latter one.

In early approaches, people normally just apply some kind of machine learning method to music generation and try to capture the music style from the training data. Walter Schulze and Brink van der Merwe [1] build the hidden Markov models to learn from a certain music style and calculate the parameters for Markov chains from the mathematical representation. Chun-Chi J. Chen and Risto Miikkulainen's work [2] is one of the earliest paper which introduces neural network method to music generation. They use a multi-layer neural network to build the melody part but pay little attention to other things such as chords.

For recent methods, time-series models are usually adopted, and many of them deploy more dedicated structures to build and represent the various time dependencies in music structure. Briot, Hadjeres, and Pachet 2017 [3] publish a detailed survey study of music generation on deep learning methods, which is a good reference to start. Many methods use LSTM-style models to construct the pipeline. Huang A, Wu R.'s work [4] provides a simple but effective baseline for end-to-end learning and generative method. Kalingeri V, Grandhe S. [5] further experiment with more variations of LSTM and also adopt convolution layers. There are also more novel ideas to introduce other kinds of deep learning methods for this field. Boulanger-Lewandowski, Bengio, and Vincent [6] bring up RNN-RBM for music generation, a recurrent temporal restricted Boltz-Mann machine to model symbolic sequences of polyphonic music in a piano-roll representation. Dong H W, Hsiao W Y, Yang L C, et al. [7] introduces generative adversarial networks (GANs) and propose a multi-track music generation model called MuseGAN. They also use convolution in generators and discriminators.

\section{Methods}
\label{headings}
\subsection{Data Representation}
\begin{figure}[h]
    \centering
    \includegraphics[scale=0.4]{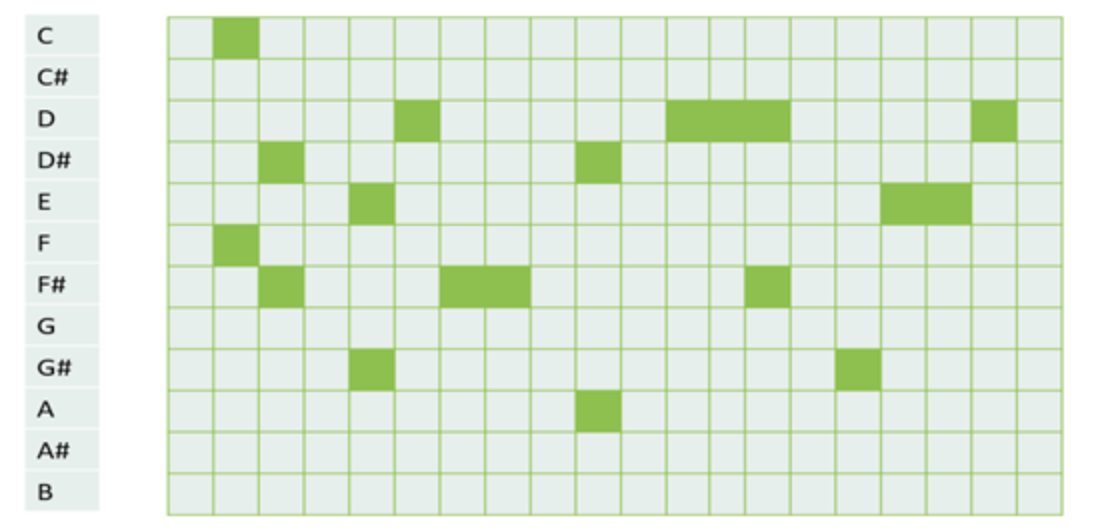}
    \caption{Pianoroll Representation}
    \label{fig:piano_roll}
\end{figure}

There are many ways to represent music numerically. Two standard ways are Wave representation, which is widely used to represent audio data, and MIDI format, which a music-specific data format. Since the purpose of our project is to learn as much as possible, we tried both of them in our method. However, we abandoned the WAVE representation due to its bad performance. 
\begin{itemize}
    \item {\bf Wave}: Waveform Audio File Format(WAV) is a very standard format for audio data. We firstly used Fast Fourier Transform(FFT) to extract the energy spectrum and use it as the feature of the original waveform. We use a mean square loss to predict the energy spectrum and then use an Inverse Fast Fourier Transform(IFFT) to convert the energy spectrum back to the waveform.
    \item {\bf MIDI-Pianoroll}: The representation of piano roll is shown in Figure \ref{fig:piano_roll}. Music is represented in a two-dimension matrix. The y-axis represents the 128 pitches on a piano and the x-axis represents the time dimension. 
    \item {\bf MIDI-Embedding}: Embedding is a concept borrowed from NLP. Embedding is specifically used to model the chord. At first, we built a corpus for all the chords in the training method. For example, "C, F, D" represents one chord and we assign a unique index for it. During training, an embedding layer is placed between LSTM and the input layer. So in a sense, each chord has a vector representation, i.e., chord embedding.
\end{itemize}

\begin{figure}[h]
    \centering
    \includegraphics[scale=0.4]{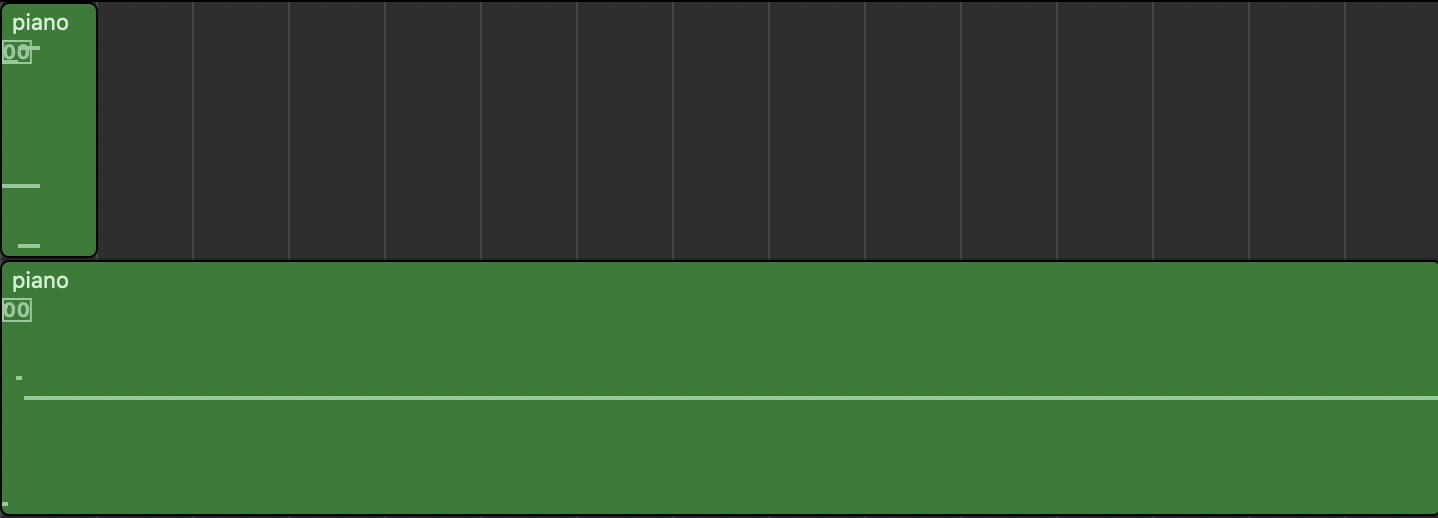}
    \caption{piano roll Examples of the Drawback of Naive Training and Generating method. Upper one represents short generation and the bottom one represents the constant generation.}
    \label{fig:set}
\end{figure}  

\subsection{Modeling the Temporal Structure}
Music is a very typical example of time-series data. Not only harmony not also coherency of a piece of music heavily depends on the time characteristic of music. Recurrent Neural Network is well-known for its capability to capture the time-series dependence so, in our project, we relied on the different architecture of RNN to generate the music.
\begin{itemize}
    \item {\bf Simple LSTM Model}: The Long Short Term Memory (LSTM) structure uses gates method to store the state information of long term previous information, which makes it strong in reserve the information in the previous timestamp. In this baseline model, we just simply use two LSTM cells to deal with the time-series information and then employ one linear layer followed by the output.

    \begin{figure}[h]
    \centering
    \includegraphics[scale=2]{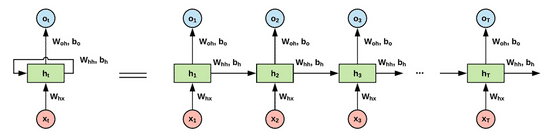}
    \caption{LSTM Structure}
    \label{fig:set}
    \end{figure}    
    
    \item {\bf Simple LSTM Encoder-Decoder Model}: Based on the previous one, this seq-2-seq model consist of two LSTM models. As shown in Figure \ref{fig:en_de}, it specifies the input and output section by using the encoder to encode the input and then using the decoder to decode the data respectively.
    
    \begin{figure}[h]
    \centering
    \includegraphics[scale=0.4]{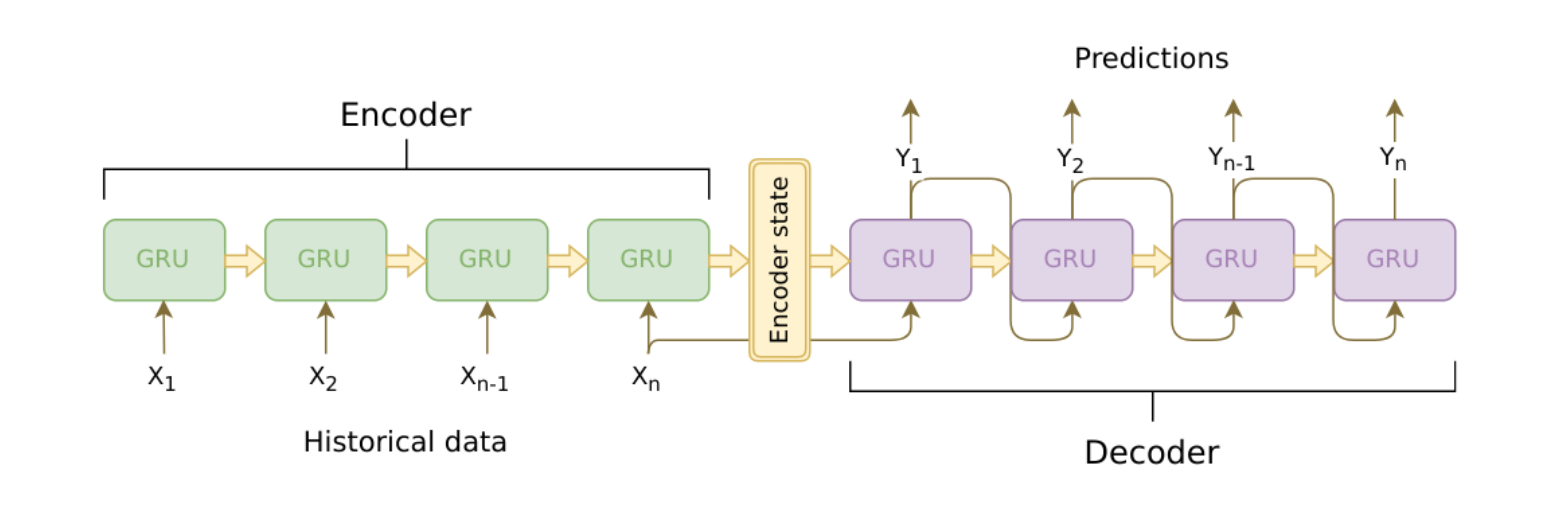}
    \caption{LSTM Encoder-Decoder Structure}
    \label{fig:en_de}
    \end{figure}    

    \item {\bf Attention-Based-LSTM Encoder-Decoder Model}: In this setting, besides the existing encoder-decoder structure, we further add the attention module. Its structure is shown in Figure \ref{fig:attention}. The attention mechanism weights the output of the encoder and passes them to the decoder, which provides more context information for the decoder section and shows the relationship between decoder and encoder in each timestamp.
    
    \begin{figure}[h]
    \centering
    \includegraphics[scale=0.9]{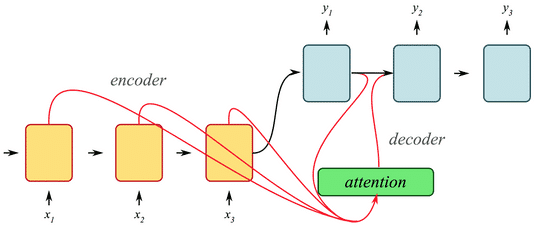}
    \caption{Encoder-Decoder with Attention}
    \label{fig:attention}
    \end{figure}    
    
    \item {\bf CNN+Attention-Based-LSTM Encoder-Decoder Model}: To fully make use of the inter-dependence between timestamp and notes. We used two convolutional layers in our model to help the network learn the relationship between time and notes each of them is a 1-D convolution on the data. For the timestamp convolution, we set the kernel size as 10 to match the concept of bars in the music. As for note convolution, we set kernel size as 11 to match the note number in each octave. We find that such convolution structure could improve the performance of the Pianoroll representation for it conform to the theory of music, which also prove the importance of field knowledge in this project. And the rest of the network is an encoder and decoder structure with attention mechanism which is identical to what we did in the embedding method.
\end{itemize}

\subsection{Modeling the Dual-track Inter-dependency}
We propose a {\bf innovative method} for generating MIDI-based music, called dual-track structure. The intuition is that we want to use the field knowledge that classical piano music is played by both hands. Different hand takes different parts of music. More specifically, the main melody is played by the right hand, and the left-hand handle the low end.
\begin{itemize}
    \item {\bf Dual-Track}: The pipeline is shown in Figure \ref{fig:dual-track}. In this setting, we first use a generator to generate the right-hand part. Then, a Multi-layer Perceptron is deployed to predict the corresponding left-hand part. Finally, we combine the output from these two parts as the final output.
    
    \begin{figure}[h]
    \centering
    \includegraphics[scale=1.2]{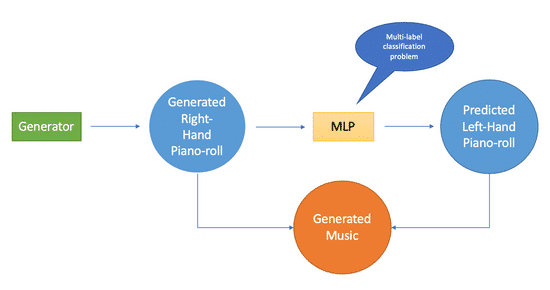}
    \caption{LSTM Encoder-Decoder Model}
    \label{fig:dual-track}
    \end{figure}    

\end{itemize}

\section{Experiments}
\label{others}

\subsection{Dataset}
We used two datasets for training and testing. The first one is a collection of 8 classical piano music, which are carefully selected from this \hyperlink{http://www.piano-midi.de/midi_files.htm}{link} and all have similar rhythms. The second one is \hyperlink{http://abc.sourceforge.net/NMD/}{Nottingham dataset}, which contains over 100 music.

\subsection{Model Settings}
Since our model is LSTM-based, the dimension of our input is $(TimeDim \times Batch Size \times Embedding Size)$, to mostly leverage our field knowledge in music, we 4 bars to predict the next 4 bars, which means, under the assumption that there are 3/4 time signature and 24 timestamps per beat, each bar contains 3 * 24 = 72 timestamps. And the embedding size is 200 for Embedding representation, 128 for piano roll representation (Each piano only contains a maximum 128 pitches.)
\subsection{Training and Generation Paradigm }
When training our model, we made a lot of mistakes. The main two challenges are 1) Generated music is short. In other words, the generator always generates zeros after a small number of generation rounds. 2) The generated music is mostly unvarying. That is to say, a generator tends to generate the same chord all the time. After researching, we used two {\bf innovative techniques} from machine translation to train and generate our music samples.
\begin{itemize}
    \item {\bf Teaching Force Rate}: When training an LSTM model, regardless of Simple LSTM or decoder of the Encoder-Decoder model, we introduce a teaching force rate $p$ to change the training input dynamically. For example, in the standard training method, the input for a model is the true input from the training set. But if we set $p=0.2$, then there's a 20\% chance that the predicted output of the model in the last timestamp would be the input for this timestamp. In this way, the model could be capable of generating longer music. 
    \item {\bf Randomization on Generation}: Due to the observations that our model usually generates some constant pitches, i.e., only one chord is generated by the generator. We introduced the randomization method to encounter this problem. Unlike machine learning translation problems, in the music generation task, we don't have a clear and deterministic target output. Oppositely, we wish our model could generate different music every time based on the manifold it learned. The first thing we tried is Top-K sampling. At each time of timestamp, we do not greedily pick the label with the highest probability at that time but randomly choose among the top k labels that have the highest probability. Such techniques indeed add some randomness in our generated music, but it also makes the model unstable and the music generated by it turned to oscillate too much. Our next method is to use a Gumbel noise technique, which is to add a random noise under Gumbel Distribution upon the output probability distribution before we soft-max and pick the label. This method is much stable than random sampling and we could use the parameters of this Gumbel distribution as hyper-parameters to adjust the magnitude of randomness we introduced.
    
\end{itemize}
\subsection{Objective Metrics for Evaluation}
\label{metric}
We adopt two paradigms for evaluation. The first one is a manual evaluation by looking at the generated MIDI output. This approach is intended to choose our best model for each data representation method. More specifically, we consider aspects of training convergence, over-fitting, generated length and harmony of the music. 

After picking our best model, we then use the following quantitative metrics to evaluate the results and compare them with the industrial state-of-the-art methods and true music.
\begin{itemize}
    \item {\bf UPC}: Number of used pitch classes per bar (from 0 to 12).
    \item {\bf QN}: Ratio of “qualified” notes in percent. We consider a note no shorter than three time steps (i.e. a 32-th note) as a qualified
note. {\bf QN} shows if the music is overly fragmented.
    \item {\bf US}: User study ratings. We invited in total 10 people to evaluate how natural is the music (natural rate from 1 to 5).
\end{itemize}

\section{Evaluation}
Since music evaluation is usually subjective, it's hard to do a quantitative analysis for music generation. As stated in 4.3, we used two different methods to evaluate our models. At first, we manually look at the generated MIDI output and reason the performance of our model. Based on the results, we pick the best two models with the highest performance for MIDI-Embedding based model and the MIDI-Pianoroll respectively. Then we quantitatively evaluate the two models using the metrics in Section \ref{metric} and compare with the industrial state-of-the-art, i.e., MuseGAN and the real music.

\subsection{Manual Evaluation}
As mentioned above, we evaluated the samples of our model manually to make an initial evaluation before using a quantitative metric, because only depending on the quantitative metrics is not comprehensive and informative sufficiently for music evaluation. During the evaluation, we mainly focus on these three questions: \\

(1) Is the sample valid music or just a repeat of a single chord? \\
(2) Does the sample last a long time or end quickly after several timestamps? \\
(3) Does the sample have creativity or just the repeat of some clip of original music? \\

These three questions are considered to be the essential problems that we need to focus on. They also conform to the common way to evaluate artificial music. Note that for question 3, it is hard to evaluate using a quantitative metric, which is why some manual evaluation is necessary for our experiments.

We evaluated 10 samples of music generated by the same model, and then answer the questions above for this model. Because these three questions are quite straight forward and obvious, we use a discrete variable to answer these three questions instead of a score. In our evaluation, we almost shared the same judgment on each question for the answers are obvious for all the samples. The evaluation results are shown below:

\begin{table}[H]
    \centering
    \begin{tabular}{c|c|c|c}
    \hline
    {\bf Models} &  {\bf Valid} & {\bf Length} & {\bf Creativity}\\
    \hline
         Waveform & No & - & - \\
         Embedding + En-De. & Yes & Short & - \\
         Embedding + LSTM & Yes & Long & Completely Overfit \\ 
         Embedding Atten-LSTM En-De.& Long & Long & \bf Yes \\
         Pianoroll + LSTM & Yes & Short & - \\
         Pianoroll + En-De. & Yes & Long & Slightly Overfit \\ 
         Pianoroll CNN+Atten-LSTM En-De.& Yes & Long & \bf Yes \\
         \hline
    \end{tabular}
    \caption{Manual Analysis of Samples of Our Models}
    \label{tab:manual-metrics}
\end{table}

Table \ref{tab:manual-metrics} indicates the progress we made in our model and also the timeline we implemented them. As it shows, our model didn't work well at the beginning and we analyzed why such a problem happens. Our analysis focus on the three questions we answered and what led to the failure of our model in the beginning.
\begin{itemize}
\item{\bf Failure of Waveform}: The first method we tried is waveform prediction, which is not valid because of noises. The prediction on the frequency field is hard to converge and it is in bad shape during the generating process, which will lead to obvious noises when using IFFT to transform into the time field. So, we didn't move on this track.

\item{\bf Extending the length}: During the evaluation, we find the lengths of generated music are limited. Our models tended to become saturated in several timestamps and began to generate empty chords. A technique to address this is the teaching force, which we have mentioned above. Because such a technique may jeopardize the converge process, we first set this probability to 0 and then increase it when we were training in higher epochs. We find this fairly helpful for training to make the models generating longer music pieces.

\item{\bf Toward Pianoroll}: In experiments, we explored two different ways of representing music and we found that they can lead to different model performances. The Piano roll method uses a non-one-hot vector as the label, while the Embedding method using an embedding layer after the one-hot label. In training, we found that compared with the embedding method, the Piano roll method is hard to converge and produce shorter chord sequences. The reason is that the non-one-hot label in Piano roll is not a good representation of music intuitively. Some non-one-hot labels are close in the piano keyboard which makes them have smaller numeric distance, but in reality, they sound completely different. So, to help the network have a sense of circular key in the piano, we deployed a CNN layer to learn the inter-dependence between the chords, which led to a better result for the Pianoroll method.

\item{\bf LSTM and En. and De.}: Another important finding in the experiments is the over-fitting problem in LSTM. LSTM is a strong predictor that could fully exploit the time information in the data, but it is easy to become over-fitting in practice. In our experiments, it could fully memorize the small dataset and repeat some clips of the original music as shown in Figure \ref{fig:overfit} , which makes it had a bad performance. Although we tried to introduce randomness and regularization to it, the generated sequences still tended to have an identical output with the original music. On the other hand, as a classical generator, the Encoder-Decoder structure has a bottleneck structure inside, which will force the network to learn the manifold of the music and generate music based on this manifold. So, the Encoder-Decoder structure could generate music with more creativity and this is why we selected it as the based generator in our final model.

\begin{figure}[h]
    \centering
    \includegraphics[scale=0.6]{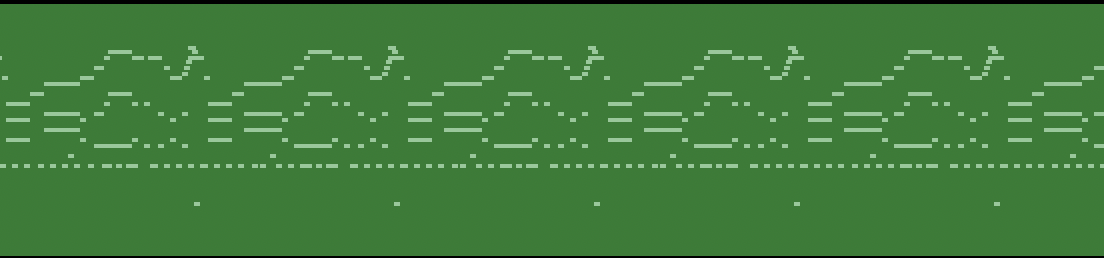}
    \caption{A overfit sample generated by LSTM}
    \label{fig:overfit}
    \end{figure}  
\end{itemize}

Based on this founding, we refined our model by using an Encoder-Decoder structure along with the attenthion module, and then use quantitative evaluation to compare the results with MuseGAN and true music.

\subsection{Quantitative Evaluation}
Table \ref{tab:quant-metrics} shows the quantitative analysis of our models. {\bf UPC} measures the variability of our generated music. Higher {\bf UPC} means that the model prefers to generate more complex chords. It is good in the sense that the music generated is more complex but this does not guarantee it to be coherent and harmonious. Therefore, {\bf QN} is introduced to measure the quality of prediction. Under this policy, we find our Pianoroll CNN+Atten-LSTM En-De has the highest {\bf QN}. This is because CNN could successfully capture the local structure of the input data. Our Pianoroll Embedding Atten-LSTM Encoder-Decoder model has very high {\bf UPC} because it directly predicts the chords rather than pitches. So Embedding representation is good at handling chords in the music. The score of MuseGAN is borrowed from the original paper. Because MuseGAN handles the Multi-track problem and the architecture is much more complex than ours, it's reasonable to find that both {\bf QN} and {\bf UPC} are smaller.

In terms of user study, we present our samples to 10 people and let them rate how natural the music is. Among our models, Embedding got the highest score due to it has high {\bf UPC}. MuseGAN has a higher score because it generates not only piano music but also drums and bass.
\begin{table}[H]
    \centering
    \begin{tabular}{c|c|c|c}
    \hline
    {\bf Models/Metrics} &  {\bf UPC} & {\bf QN} & {\bf US}\\
    \hline
         True Music & 9.83& 98.7\% & 3.8\\ 
         MuseGAN & 4.57 & 64\% & {\bf 3.16}\\ 
         Pianoroll CNN+Atten-LSTM En-De.&2.35 & {\bf 91.2}\%& 2.4 \\ 
         Embedding Atten-LSTM En-De.& {\bf 7.79} & 90\% & 2.7 \\
         \hline
    \end{tabular}
    \caption{Quantitative Analysis of Our Models with MuseGAN and True Music}
    \label{tab:quant-metrics}
\end{table}
\section{Conclusion}

Our wave model showed the worst performance and our dual-track embedding model using the attention-encoder-decoder generator outperforms the other models and achieves the best performance. In conclusion, we find that the Embedding method generally performs better than the piano roll representation. But once we add CNN structure to leverage the local interdependence for pianoroll-based models, they also show comparable performances.
\section{Future Work}
In this work, we only trained our models on the small-scale dataset. We believe there is a good chance for them to perform better when trained and tuned on a larger dataset. Also, there are more methods that are also good at generation tasks, such as GAN and VAE. They should be a good start to try out for further work.
\section*{References}

\medskip

\small

[1] Schulze W, Van Der Merwe B. Music generation with Markov models[J]. IEEE MultiMedia, 2010 (3): 78-85.

[2] Chun-Chi J. Chen and Risto Miikkulainen. Creating melodies with evolving recurrent neural networks. Proceedings of the 2001 International Joint Conference on Neural Networks, 2001.

[3] Briot J P, Hadjeres G, Pachet F. Deep learning techniques for music generation-a survey[J]. arXiv preprint arXiv:1709.01620, 2017.

[4] Huang A, Wu R. Deep learning for music[J]. arXiv preprint arXiv:1606.04930, 2016.

[5] Kalingeri V, Grandhe S. Music generation with deep learning[J]. arXiv preprint arXiv:1612.04928, 2016.

[6] Boulanger-Lewandowski N, Bengio Y, Vincent P. Modeling temporal dependencies in high-dimensional sequences: Application to polyphonic music generation and transcription[J]. arXiv preprint arXiv:1206.6392, 2012.

[7] Dong H W, Hsiao W Y, Yang L C, et al. MuseGAN: Multi-track sequential generative adversarial networks for symbolic music generation and accompaniment[C]//Thirty-Second AAAI Conference on Artificial Intelligence. 2018.

\end{document}